\documentstyle[12pt,aaspp4]{article}
\begin{document}

\slugcomment{Astronomical Journal, in press (Jan 2001)}

\title{High-Redshift Quasars Found in Sloan Digital
Sky Survey Commissioning Data IV:
Luminosity Function from the Fall Equatorial Stripe Sample$^1$}

\author{Xiaohui Fan\altaffilmark{\ref{Princeton},\ref{IAS}},
Michael A. Strauss\altaffilmark{\ref{Princeton}},
Donald P. Schneider\altaffilmark{\ref{PennState}},
James E. Gunn\altaffilmark{\ref{Princeton}},
Robert H. Lupton\altaffilmark{\ref{Princeton}},
Robert H. Becker\altaffilmark{\ref{UCDavis},\ref{IGPP}},
Marc Davis\altaffilmark{\ref{Berkeley}},
Jeffrey A. Newman\altaffilmark{\ref{Berkeley}},
Gordon T. Richards\altaffilmark{\ref{Chicago},\ref{PennState}},
Richard L. White\altaffilmark{\ref{STScI}},
John E. Anderson, Jr.\altaffilmark{\ref{Fermilab}},
James Annis\altaffilmark{\ref{Fermilab}},
Neta A. Bahcall\altaffilmark{\ref{Princeton}},
Robert J. Brunner\altaffilmark{\ref{Caltech}},
Istvan Csabai\altaffilmark{\ref{JHU}},
Mamoru Doi\altaffilmark{\ref{UTokyo}},
G. S. Hennessy\altaffilmark{\ref{USNO}},
Robert B. Hindsley\altaffilmark{\ref{NRL}},
Masataka Fukugita\altaffilmark{\ref{CosmicRay},\ref{IAS}},
Peter Z. Kunszt\altaffilmark{\ref{JHU}},
\v{Z}eljko Ivezi\'{c}\altaffilmark{\ref{Princeton}},
Gillian R. Knapp\altaffilmark{\ref{Princeton}},
Timothy A. McKay\altaffilmark{\ref{Michigan}},
Jeffrey A. Munn\altaffilmark{\ref{Flagstaff}},
Jeffrey R. Pier\altaffilmark{\ref{Flagstaff}},
Alexander S. Szalay\altaffilmark{\ref{JHU}},
and Donald G. York\altaffilmark{\ref{Chicago}}
}

\altaffiltext{1}{Based on observations obtained with the
Sloan Digital Sky Survey, 
and with the Apache Point Observatory
3.5-meter telescope, 
which is owned and operated by the Astrophysical Research Consortium,
by the W. M. Keck Observatory, which is operated as a scientific
partnership among the California Institute of Technology,
the University of California, and NASA,
and was made possible by the generous financial support of the W. M. Keck
Foundation,
and with the Hobby-Eberly Telescope,
which is a joint project of the University of Texas at Austin,
the Pennsylvania State University, Stanford University,
Ludwig-Maximillians-Universit\"at M\"unchen, and Georg-August-Universit\"at
G\"ottingen.
}
\newcounter{address}
\setcounter{address}{2}
\altaffiltext{\theaddress}{Princeton University Observatory, Princeton,
NJ 08544
\label{Princeton}}
\addtocounter{address}{1}
\altaffiltext{\theaddress}{Institute for Advanced Study, Olden Lane,
Princeton, NJ 08540
\label{IAS}}
\addtocounter{address}{1}
\altaffiltext{\theaddress}{Department of Astronomy and Astrophysics,
The Pennsylvania State University,
University Park, PA 16802
\label{PennState}}
\addtocounter{address}{1}
\altaffiltext{\theaddress}{Physics Dept., University of California, Davis, CA 95616
\label{UCDavis}}
\addtocounter{address}{1}
\altaffiltext{\theaddress}{IGPP/Lawrence Livermore National Laboratory
\label{IGPP}}
\addtocounter{address}{1}
\altaffiltext{\theaddress}{Department of Astronomy, University of California, Berkeley, CA 94720-3411
\label{Berkeley}}
\addtocounter{address}{1}
\altaffiltext{\theaddress}{University of Chicago, Astronomy \& Astrophysics Center and the Enrico Fermi Institute, Chicago, IL 60637
\label{Chicago}}
\addtocounter{address}{1}
\altaffiltext{\theaddress}{Space Telescope Science Institute, Baltimore, MD 21218
\label{STScI}}
\addtocounter{address}{1}
\altaffiltext{\theaddress}{Fermi National Accelerator Laboratory, P.O. Box 500,
Batavia, IL 60510
\label{Fermilab}}
\addtocounter{address}{1}
\altaffiltext{\theaddress}{
Department of Astronomy, California Institute of Technology,
Pasadena, CA 91125
\label{Caltech}}
\addtocounter{address}{1}
\addtocounter{address}{1}
\altaffiltext{\theaddress}{
Department of Physics and Astronomy, The Johns Hopkins University, Baltimore, MD 21218
\label{JHU}}
\addtocounter{address}{1}
\altaffiltext{\theaddress}{Department of Astronomy and Research Center for the
 Early Universe,
        School of Science, University of Tokyo, Hongo, Bunkyo,
Tokyo, 113-0033 Japan
\label{UTokyo}}
\addtocounter{address}{1}
\altaffiltext{\theaddress}{Institute for Cosmic Ray Research, University of
Tokyo, Midori, Tanashi, Tokyo 188-8502, Japan
\label{CosmicRay}}
\addtocounter{address}{1}
\altaffiltext{\theaddress}{U.S. Naval Observatory,
3450 Massachusetts Ave., NW,
Washington, DC  20392-5420
\label{USNO}}
\addtocounter{address}{1}
\altaffiltext{\theaddress}{Remote Sensing Division, Code 7215, Naval
  Research Laboratory, Washington, DC 20375
\label{NRL}}
\addtocounter{address}{1}
\altaffiltext{\theaddress}{University of Michigan, Department of Physics,
        500 East University, Ann Arbor, MI 48109
\label{Michigan}}
\addtocounter{address}{1}
\altaffiltext{\theaddress}{U.S. Naval Observatory, Flagstaff Station,
P.O. Box 1149,
Flagstaff, AZ  86002-1149
\label{Flagstaff}}

\begin{abstract}
This is the fourth paper in a series aimed at
finding high-redshift quasars from five-color ($u'g'r'i'z'$)
imaging data taken along the Celestial Equator by
the Sloan Digital Sky Survey (SDSS)
during its commissioning phase.
In this paper, we use the color-selected sample of 
39 luminous high-redshift quasars presented
in Paper III to derive the evolution of the quasar luminosity function
over the range of $3.6 < z < 5.0$, and $-27.5 < M_{1450} < -25.5$  
($\Omega=1$, $H_{0}$ = 50 km s$^{-1}$ Mpc$^{-1}$).

We use the selection function derived in Paper III to correct
for sample incompleteness.
The luminosity function is estimated using three different methods:
(1) the $1/V_{a}$ estimator;
(2) a maximum likelihood solution,  assuming 
that the density of quasars depends 
exponentially on redshift and as a power law in luminosity,
and (3) Lynden-Bell's non-parametric $C^{-}$ estimator. 
All three methods give consistent results.
The luminous quasar density decreases by a factor of $\sim 6$ from
$z=3.5$ to $z=5.0$,
consistent with the decline seen from several previous optical surveys 
at $z<4.5$.
The luminosity function follows
$\psi(L) \propto L^{-2.5}$ for $z\sim 4$ at the bright end, 
significantly flatter than the bright end luminosity function
$\psi(L) \propto L^{-3.5}$ found in previous studies
for $z<3$, suggesting that the shape of the quasar luminosity
function evolves with redshift as well, and that the
quasar evolution from $z=2$ to 5 cannot be described as
 pure luminosity evolution.
Possible selection biases and the effect
of dust extinction on the redshift evolution of the quasar density
are also discussed.

\end{abstract}

\section{Introduction}

Soon after quasars were first discovered, \cite{Schmidt68} found
them to evolve strongly with redshift.
In the following 25 years, a consensus has been reached that 
the number density of optical quasars peaks at $z\sim 2.5 - 3$, and declines towards
both lower and higher redshifts.  The luminosity function of optical
quasars at $z<2.5$ has been well studied through UV-excess and
slitless spectroscopic surveys
(e.g., \cite{Boyle88}, \cite{Hewett91}, \cite{Pei95}, \cite{2dF}).
At the high-redshift end, both multicolor surveys (\cite{WHO}, WHO hereafter; 
\cite{DPOSS}, DPOSS hereafter)
and grism surveys (\cite{SSG}, SSG here after) 
have shown that the number density
of quasars declines rapidly from $z \sim 3$ to $4.3$.
The sample sizes for these surveys are relatively small,
so the rate of this decline and the shape of the high-redshift quasar
luminosity function are not nearly as well-constrained as at low redshift.
The quasar luminosity function provides important constraints
for models of quasar evolution (e.g., \cite{ER88}, \cite{Turner91}, \cite{HL98}, \cite{HH00}), and
the nature of the high-redshift UV ionizing field (\cite{Madau99}).

In Paper III (\cite{Paper3}), we presented a sample
of 39 quasars at $z>3.6$ and $i^* \lesssim 20$, selected from 182 deg$^2$
of Fall Equatorial Stripe multicolor imaging data taken
during the commissioning phase of the Sloan Digital Sky Survey (SDSS,
\cite{York00}, see also \cite{F96}, \cite{Gunnetal}).
The selection completeness function of this multicolor
sample is calculated based on model quasar colors.
In this paper, we derive the high-redshift quasar luminosity function
based on the quasar sample and selection function presented in Paper III.
In \S2, we use three different statistical approaches to
find the luminosity function: a $1/V_{a}$
estimator; a maximum likelihood solution; 
and Lynden-Bell's non-parametric $C^-$ estimator (\cite{LB71}).
In \S3, we discuss the rate of decline of the quasar spatial density towards
high redshift and the shape of the quasar luminosity function based on
the results of \S2, and compare these new results with those of previous
studies.
In \S4, we discuss  possible selection biases on
the evolution of the luminosity function
and  theoretical implications of our results. 

\section{Derivation of Luminosity Function}

The differential quasar luminosity function, $\Psi(L,z)$, 
is defined as the number of quasars per unit comoving volume,
per unit luminosity as a function of luminosity and redshift.
It can also be written as $\Psi(M,z)$, 
the number of quasars per unit comoving volume,
per unit absolute magnitude:
\begin{equation}
\Psi(M,z) = \Psi(L,z) \left|\frac{dL}{dM}\right| = 0.92 L \Psi(L,z).  
\end{equation}
Throughout the paper, we use $\Psi(L,z)$ for the mathematical
derivations, while we present our final results in terms of the more
familiar $\Psi(M,z)$.
The quasar luminosity function can also be presented in
the cumulative form: $\Phi(L,z) \equiv \int_L^{\infty} \Psi(L',z) dL'$.

In general, the redshift and luminosity dependences of
the quasar luminosity function are not separable.
In limiting cases, $\Psi(L,z)$ can be represented either in the form of 
pure density evolution:
\begin{equation}
\Psi(L,z) = \psi(L) \rho(z),
\end{equation}
where the luminosity function remains the same shape, but with different
normalization at different redshift; or in the form of 
 pure luminosity evolution
\begin{equation}
\Psi(L,z) = \Psi[L/g(z)]/g(z),
\end{equation}
where $g(z)$, the characteristic luminosity,
is a function of redshift while the normalization of the luminosity function
is a constant.
It might be better presented by a combination of the two cases above
(e.g., \cite{KK88}):
\begin{equation}
\Psi(L,z) = \rho(z) \Psi[L/g(z)]/g(z).
\end{equation}

For $z\lesssim 3$, pure luminosity evolution provides a good fit to
the quasar luminosity function (e.g. \cite{PG78}, \cite{2dF}).
At $z>3$, existing quasar samples are still too small 
to distinguish these models.
In general, the evolution can be more
complicated. For example, the luminosity function at low and high redshifts
may be determined by very different physics and have totally different
shapes.
Note that if the luminosity function is a power law, $\Psi(L,z)  \propto
L^\beta$, where $\beta$ does not depend on $z$,
there is no characteristic luminosity, 
and pure luminosity and pure density evolution cannot be distinguished.
 
The determination of the high-redshift quasar luminosity function from a 
flux-limited sample faces several  difficulties:
(1) the selection function of a multicolor survey is a strong function of luminosity,
redshift and the shape of the quasar spectral energy distribution 
(SED, see Paper III); the selection bias has
to be corrected for carefully.
(2) The high-redshift quasar samples are typically small
(the largest complete sample of quasars  at $z>4$
prior to this work has 10 objects).
The small sample size makes it difficult to 
parameterize the luminosity function, and the result is also strongly
affected by the choice of binning. 
(3) As the redshift increases, an optical survey samples different
parts of the quasar's intrinsic spectrum. 
When comparing results from low and high redshift samples, the K-correction
is large and quite uncertain, due to the lack of statistics on the
continuum shape between the rest-frame UV and optical region 
of high-redshift quasars.

The luminosity function can be derived using either non-parametric
or parametric methods. 
The most commonly used  non-parametric method is  
the $1/V_a$ estimator (e.g. \cite{Avni80}). 
We present the $1/V_a$ results in \S2.1.
In \S2.2, we generalize the test of correlation between luminosity 
and redshift distributions described in \cite{EP92} and \cite{MP99}, 
by including the selection function.
We show that for the limited range of redshift and luminosity our
survey covers, the two distributions are not
correlated.
Thus in \S2.3, we assume that the
luminosity and redshift dependences are separable, and
that the quasar density follows a power law in luminosity 
and depends on redshift exponentially. 
The maximum likelihood solutions are derived
following the method of \cite{Marshall85},
where each quasar is treated as a $\delta$ function in parameter space, and
the results are not subject to binning. 
When the density and luminosity evolution can be separated,
one can also use the Lynden-Bell's $C^-$ estimator (\cite{LB71})  to calculate
the non-parametric marginal distributions of the cumulative luminosity function
along the luminosity and redshift directions.
In Appendix A, we show how to generalize the $C^-$ estimator to include
the survey selection function.
The luminosity function estimated using the $C^-$ estimator is presented
in \S2.4.

The results throughout this paper are presented for two cosmological
models: (1) a model with $\Omega = 1$ and $H_0 = 50$ km s$^{-1}$ Mpc$^{-1}$,
which we refer to as the $\Omega =1$ model. Most of the previous
studies of the quasar luminosity function have used this cosmology.
(2) a $\Lambda$-dominated flat model with $\Omega = 0.35$, $\Lambda$=0.65,
and $H_0 = 65$ km s$^{-1}$ Mpc$^{-1}$ (\cite{OS95}, 
\cite{KT95}, \cite{Bahcall99}), which we refer to as
the $\Lambda$-model.
The luminosity function is presented in term of $M_{1450}$, the
absolute AB magnitude of the quasar continuum in the rest-frame
at 1450 \AA. Assuming a power law $f_{\nu} \propto \nu^{\alpha}$,
Schmidt et al. (1995) find:
\begin{equation}
M_{B} = M_{1450} + 2.5\alpha\log(4400/1450) + 0.12,
\end{equation}
where the effective wavelength of the Kron-Cousins B band is 4400 \AA, 
and the factor 0.12 comes from the zero point difference between the
AB and Vega-based magnitude systems for quasar-like spectra.

\subsection{1/${V_a}$ estimate}

We first calculate the space density of quasars in different luminosity and
redshift bins using the $1/V_a$ method 
following the discussions in WHO and DPOSS.
The available volume $V_a$ is defined for a quasar of given
luminosity and SED as the comoving
volume over which the quasar could have been detected by the survey.
In the presence of a selection function, the available volume
is weighted by the selection probability.
If one is to determine the luminosity in 
luminosity and redshift bins
$\Delta z, \Delta L$,
the available volume for a quasar with given $L$, $z$ and SED is:
\begin{equation}
V_a  = \int_{\Delta z} p(L,z,\mbox{SED})\frac{dV}{dz} dz,
\end{equation}
where $p(L,z,\mbox{SED})$ is the selection probability of the quasar
as a function of redshift, luminosity and the SED.
The luminosity function and its statistical uncertainty can be estimated
as:
\begin{equation}
\Psi(\langle L \rangle, \langle z \rangle) = \sum_{i} \frac{1}{V_a^i \Delta L}
\end{equation}
\begin{equation}
\sigma(\Psi) = \left[\sum_{i} \left(\frac{1}{V_a^i \Delta L} \right)^2\right]^{1/2}
\end{equation}
where $\langle L \rangle$ and $\langle z \rangle$ are the
average luminosity and redshift over the bin, the sum is over all the objects
in the bin, and the available volume $V_a^i$ is calculated for each object.

In Paper III, two flavors of selection probability are calculated:  
(1) the probability for a given quasar SED shape, $p[L, z,\alpha, \mbox{EW
(Ly$\alpha$+NV)}]$, 
where $\alpha$ is the slope of the power law continuum and EW(Ly$\alpha$+NV)
is the rest-frame equivalent width of the Ly$\alpha$+NV emission line.
These two parameters characterize the shape of the SED.
The selection probability of each quasar in the sample is listed in Table 4
of Paper III.
(2) The average selection probability $p(L, z)$ (\S 5.3 in Paper III),
averaged over the intrinsic distribution of $\alpha$ and EW.
In Paper III, we assume that this distribution is the 
product of Gaussians in $\alpha$ and $\rm EW (Ly\alpha+NV)$,
and find that their mean and standard deviations are:
$\alpha = -0.79 \pm 0.34$ and
$\rm EW (Ly\alpha+NV) = 69.0 \pm 18.3$ \AA.
In Figure 1, we present the $1/V_a$ results of the quasar luminosity function
using both flavors of selection probability under the $\Omega=1$ and the
$\Lambda$-model cosmologies.
Figure 1(a) and (b) are for the $\Omega=1$  model.
In Figure 1(a), the  density is calculated using the
average selection function $p(L,z)$,
while in Figure 1(b), the density is calculated using
the selection probability of each quasar for
its SED type, $p[L, z,\alpha, \mbox{EW(Ly$\alpha$+NV)}]$.
Similarly, Figure 1(c) and (d) are for the $\Lambda$-model.
In each case, we divide the sample of 39 quasars in 3 redshift bins:
$3.6 < z < 3.9$ (18 quasars), $3.9 < z < 4.4$ (14 quasars) and $4.4 < z < 5.0$ (7 quasars).
The redshift bins are further divided into two or three luminosity bins.
There are 3 -- 6 quasars in each bin.

The different methods of treating the selection function give  
consistent results.
This can be understood as follows.
For simplicity, we ignore the dependence on EW, $z$ and $L$, and
assume that the quasar distribution only depends on $\alpha$.
For an intrinsic  distribution function $f(\alpha$) and
a selection function $p(\alpha$), the observed distribution
will be $n(\alpha) = p(\alpha)f(\alpha)$.
In the discrete case, $n(\alpha) = \sum_i \delta(\alpha - \alpha_i)$.
The effective density of quasars  
can be estimated as:
\begin{eqnarray}
\rho_{\rm eff}  & =  & \frac{1}{V_a} \int f(\alpha)\ d\alpha \nonumber \\
      & = & \frac{1}{V_a} \int \frac{p(\alpha)f(\alpha)\ d\alpha}{p(\alpha)} \nonumber \\
      & =  & \frac{1}{V_a} \int \frac{n(\alpha)\ d\alpha}{p(\alpha)} \nonumber \\
      & =  & \frac{1}{V_a} \sum_i \frac{1}{p(\alpha_i)}. 
\end{eqnarray}
Alternatively, if we define the average selection probability:
\begin{equation}
\overline{p} = \frac{\int p(\alpha) f(\alpha)\ d\alpha}{\int 
 f(\alpha) d\alpha}, 
\end{equation}
then  the density can also be estimated as:
\begin{equation}
\rho_{\rm eff} = \frac{\int p(\alpha) f(\alpha)\ d\alpha}{\overline{p} V_a} = 
\frac{N}{\overline{p} V_a},
\end{equation} 
where $N$ is the total number of objects observed.
Therefore, if, as we have assumed, the SED shape is not a 
function of redshift or luminosity,
the two approaches only differ in the different weightings  applied 
to the observed $n(\alpha)$ distribution;  in Eq. (11),
each object is given the same weight and 
normalized by the {\em average} selection probability.
The two methods should give consistent results if  
the sample is large enough and
all the SED shapes are properly sampled.
However, as we showed in Paper III, there are regions in the $(L, z, \alpha,
\mbox{EW})$ parameter space poorly sampled by  the selection criteria 
($p \ll 1$).
In this case, the average probability $p(L,z)$ is much better behaved, and
we will use it for most of the analyses which follow.


Due to the small number of quasars in each bin, the appearance of
Figure 1 changes somewhat with different binning,
but the general trends are obvious:
(1) the overall density of quasars declines rapidly with redshift.
For the $\Omega = 1$ model, 
the density of quasars at $M_{1450} \sim -26.5$ decreases from
$\sim 2 \times 10^{-8}$ Mpc$^{-3}$ mag$^{-1}$ at $z=3.75$ to 
$\sim 6 \times 10^{-9}$ Mpc$^{-3}$ mag$^{-1}$ at $z=4.75$.
(2) the luminosity function rises towards lower luminosity 
by a factor of $\sim 4$ per magnitude.
These results are very similar for the $\Lambda$-model.
At $z\sim4$, the behavior of the $\Omega=1$ and $\Lambda$ models
are quite similar, as $\Omega(z) \rightarrow 1$ when the redshift becomes 
large.

\subsection{Test of Correlation between the Luminosity and Redshift Distributions}
 
In this subsection, we test whether the intrinsic 
luminosity and redshift distributions
of the quasar sample are correlated or whether
they can be treated separately
(as in the case of pure density evolution, Eq. 2).
At low redshift, the luminosity function can be fit with
a double power law (e.g., \cite{Boyle88}).
The sample in this paper covers a limited range 
of absolute magnitude
($-27.5 < M_{1450} < -25.5$) and a limited redshift range ($3.6 < z < 5.0$).
If, in this case, the luminosity function can be fit with a single
power law  whose slope is independent of the redshift, 
the luminosity and redshift distributions of our sample are not expected 
to be correlated.
This correlation test is also crucial for the non-parametric $C^-$ estimator
in \S2.4,  which 
requires that the data be expressed in terms of two uncorrelated variables. 

Correlation tests for truncated data sets (such as 
flux-limited samples) have been developed by Efron \& Petrosian (1992) and
\cite{MP99}.
In our survey, there is essentially no bright-end truncation in
the selection (the saturation limit is $i^* \sim 14$, four magnitudes
brighter than the brightest quasar in the sample).
Thus, our data set $\{L_i, z_i\}$ is one-side truncated in luminosity.
We first consider the case in which the selection function is a step function 
and the selection boundary is sharp:
\begin{equation}
p(L, z) = \left\{ \begin{array}{ll}
		1 &\mbox{if $L > L^-(z)$} \\
		0 &\mbox{if $L < L^-(z)$}
		  \end{array}
	\right.
\end{equation}
where $L^-(z)$ is the limiting luminosity at a given redshift. 
In this case, following the discussion in
\cite{MP99}, one first defines the {\em comparable} or {associated} set for
each object $i$:
\begin{equation}
J_i = \{j: L_j > L_i, L_j^- < L_i\}
\end{equation}
consisting of all objects which are brighter than the object in question
and which would be selected by the survey if they had the same luminosity as
the object in question.
This is the same set that will be used in the
$C^-$ estimator below. 
It is the  {\em luminosity} and {\em volume}
limited subset of the sample that can be constructed for each object.
If $z$ and $L$ are independent, the rank $R_i$ of $z_i$ in the
comparable set, defined as 
\begin{equation}
 R_i = \mbox{number of $j$, such that }\hspace{0.2cm}  \{ j \in J_i; z_j < z_i \} ,
\end{equation}
should be distributed uniformly between 0 and $N_i$,  where $N_i$ is
the number of points in the comparable set for each object
(not including the object in question).
The expectation value  of $R_i$, is $E_i = (1/2)N_i$ and the
variance $V_i = (1/12) N_i^2$.  
Therefore, one can construct Kendell's $\tau$ statistic:
\begin{equation}
\tau = \frac{\sum (R_i - E_i)}{\sqrt{\sum V_i}}.
\end{equation}
For $|\tau| \lesssim 1$, the luminosity and redshift would not be correlated
at $\sim 1 \sigma$ level and could be treated independently.
 
However, the discussion above only applies to a sample with a sharp 
boundary. 
For  the quasar survey we discuss in this paper, even though the selection
criteria have a sharp cut at the limiting apparent magnitude ($i^* < 20.05$ for
$gri$ selected quasars and $i^* < 20.2$ for $riz$ selected quasars,
see Paper III), 
the selection function smoothly approaches zero at the limiting luminosity
(Figure 8 of Paper III).
There is no one-to-one correspondence between the apparent $i^*$ magnitude
and the luminosity of the quasar at a given redshift, 
due to scatter in the quasar SED shape
and photometric errors in the SDSS measurements. 
We thus generalize the correlation test to include
the selection function $p(L,z)$.
First, we define a generalized comparable set:
\begin{equation}
J_i = \{j: L_j > L_i\},
\end{equation}
which includes all the objects more luminous than the object in
question, since all might have a selection probability between 0 and 1.
We then define the total number in the set by weighting
each point $j$ in the comparable set $J_i$.
This weight is proportional 
to $p(L_i, z_j)$, the selection probability if the object $j$ 
had the same luminosity $L_i$ 
as the object in question (object $i$), and it is proportional to
the inverse of its own selection probability $p(L_j, z_j)$.
It is useful to define a quantity $T_i$:
\begin{equation}
T_i =  \sum_{j=1}^{N_i} \frac{p(L_i, z_j)}{p(L_j, z_j)}.
\end{equation}
The definition of the rank changes accordingly:
\begin{equation}
R_i = \sum_{j=1}^{N_i} \frac{p(L_i, z_j)}{p(L_j, z_j)}, \hspace{0.5cm} \mbox{if}\hspace{0.2cm} z_j < z_i.
\end{equation}
The expectation value of the distribution $R_i$ is
$E_i = (1/2)T_i$, 
and the variance 
$V_i = (1/12)(T_i)^2$.
The test statistic $\tau $ is defined in the same way as in Eq. (15).
Note that these definitions return to 
its normal form if the selection function has a sharp boundary, 
since $p(L_i,z_j) = 0$ when $L_i < L_j^-$.
In the case where the selection boundary is smooth, the ranking is
weighted by $p(L_i,z_j)/p(L_j,z_j)$. 
The selection probability always increases with 
larger luminosity and smaller photometric error.  
Therefore, since $p(L_i,z_j) \leq p(L_j,z_j)$ for $L_i > L_j$,
the weight is always between 0 and 1.
In Appendix A, we prove that the comparable set defined in this way
gives the correct $C^-$ estimator for the luminosity function
in the presence of a selection function.

We calculate the ranking statistic $\tau$ of the SDSS Equatorial
Stripe quasar sample using  the average selection function
in Paper III.
For the $\Omega = 1$ model, $\tau = -0.20$; for the $\Lambda$ model,
$\tau = -0.15$.
Therefore, for the limited redshift and luminosity range this sample
covers, the two distributions can be treated independently.
For the full quasar sample of the main SDSS survey which will cover
a much larger redshift and luminosity range, 
the possibility of luminosity evolution will need
to be introduced.
The ranking technique presented here can then be used
to determine the functional form of the luminosity evolution ($g(z)$ in 
Eq. 3) of the quasar population (e.g. Maloney \& Petrosian 1999).

\subsection{Maximum Likelihood Fits}
 
In this subsection, we model the luminosity function using
the result above that the redshift and luminosity dependences are
separable. 
Imagine dividing up the sample into small bins in $L$ and $z$,
$L_i$ and $z_j$, with sizes
$\Delta L$ and $\Delta z$. The likelihood that $n_{ij}$ quasars are found in
the ($L_i$, $z_j$) bin can be written as :
\begin{equation}
{\cal L} = \prod_{i,j} \frac{e^{-\mu_{ij}}\mu_{ij}^{n_{ij}}}{n_{ij}!},
\end{equation}
where $\mu_{ij}$ is the average number of quasars expected in the bin:
\begin{equation}
\mu_{ij} = \int \!\! \int_{\Delta L, \Delta z} \Psi(L,z) p(L,z) \frac{dV}{dz} dL\ dz,
\end{equation}
where $\Psi(L,z)$ is the differential luminosity function, and the integral is
over the bin.
The maximum likelihood solution  is obtained by minimizing the function
\begin{equation}
S = -2 \ln {\cal L},
\end{equation}
with respect to the parameters describing the luminosity function.

One could find the maximum likehood solution by fitting the $1/V_a$
results in \S2.1, but the result would be strongly affected by
the choice of binning for a small sample.
However, in the limit of an infinitesimal bin size, 
$n_{ij}$ is either 0 and 1, and the result is
not dependent on  binning.
Following \cite{Marshall85}, the likelihood function can then be written 
as:
\begin{equation}
S = - 2 \sum_i^N \ln [\Psi(L_i, z_i) p(L_i, z_i)] + 
2 \int\!\! \int \Psi(L, z) p(L, z) \frac{dV}{dz} dL\ dz,
\end{equation}
where the sum is over all quasars in the sample. 
The integral is over the entire $(L,z)$ range of the survey,
and is equals to the number of quasars expected in the survey  
for a given luminosity function.
This term provides the normalization for the likelihood function.

The likelihood function can be expressed in a slightly different
way (see also \S5.2 in Paper III):
\begin{equation}
{\cal L}' = \prod_i \frac{p_i \Psi(L_i, z_i)}{\int\!\! \int \Psi(L,z) p(M,z) \frac{dV}{dz} dL\ dz}
\end{equation}
\begin{equation}
S'=-2\ln {\cal L}' = - 2 \sum_i^N \ln [\Psi(L_i, z_i) p(M_i, z_i)] + 2N
\ln\left( \int\!\! \int \Psi(L, z) p(L, z) \frac{dV}{dz} dL\ dz \right),
\end{equation}
where $N$ is the sample size. 
As in  Eq.~(22), the integral term provides a normalization.
These two expressions are closely related.
If we define $N' = \int\! \int \Psi(L, z) p(L, z) \frac{dV}{dz} dL\ dz$,
then for any parameter $x$, it can be shown that
\begin{equation}
\frac{\partial S'}{\partial x} = \frac{\partial S}{\partial x} 
+ 2 \frac{\partial N'}{\partial x}\left (\frac{N}{N'} - 1 \right ).
\end{equation}
Thus if we constrain
the total number of objects expected
based on the maximum likelihood solution to be equal to the
total number actually observed, $N' = N$,   
Eqs. (22) and (24)
are maximized at the same parameters.
In practice, they give almost identical results.

The calculation in \S2.2 shows that the luminosity and redshift distributions
of the quasar sample are not correlated. 
We assume that the quasar density is a power law function of the 
quasar luminosity : $\psi(L) \propto L^{\beta}$, or,
expressed in terms of absolute magnitude, $\psi(M) \propto 10^{-0.4(\beta+1)M}$.
Following SSG, we further assume that the density at
a given luminosity declines exponentially with redshift
$\rho(z) \propto 10^{-Bz}$.
SSG express their results in the form of a cumulative
luminosity function
\begin{equation}
\log \Phi (z,<M_{1450}) = A - B(z-3) + C(M_{1450}+26),
\end{equation}
This is equivalent to the differential luminosity function
\begin{equation}
\Psi(z, M_{1450}) = \frac{\Psi^*}{10^{0.4[M_{1450}+26-\alpha(z-3)](\beta+1)}},
\end{equation}
where $\Psi^* = 10^{-0.4\ln10(\beta+1)A}$,
$\beta = -2.5C - 1$, and $\alpha = -2.5B/(\beta+1)$.

As discussed in \S5.2 of Paper III, the average probability $p(M,z)$
depends on the underlying distribution of quasar SED parameters
$\alpha$ and EW(Ly$\alpha$+NV).
The likelihood function (Eqs. 22 and 24) has to be maximized not only
with regard to the luminosity function, but also to the intrinsic distribution
of $\alpha$ and EW(Ly$\alpha$+NV) at the same time.
In practice, however,
the maximum likelihood solutions of the ($\alpha$, EW) distribution
and of the luminosity function 
are only weakly correlated.
The solutions can be found by first assuming the $\alpha$ and
EW distributions based on their weighted average 
({\em albeit} biased by selection effects) to
derive the first-order luminosity function and then
iterating the results. 

Using the likelihood function in Eq.~(22), we solve for the parameters of 
the cumulative luminosity function (Eq. 26). 
For the $\Omega=1$ model,
\begin{equation}
\log \Phi (z,<M_{1450}) = (-7.24 \pm 0.19) - 
(0.48 \pm 0.15 ) (z-3) + (0.63 \pm 0.10) (M_{1450}+26),
\end{equation}
where $\Phi$ is in units of Mpc$^{-3}$.
From the maximum likelihood fit, we also find
$\alpha = -0.79 \pm 0.34$ and EW(Ly$\alpha$+NV) = $69.3 \pm 18.0$ \AA,
as reported in Paper III.
The best-fit parameters in both cosmological models and in both cumulative
and differential forms are given in Table 1, and
these models are plotted as dashed lines on Figures 1 and 2.
The error bars on the parameters are estimated both by finding the parameters
that yield  $ S = S_{min} + 1$ (which gives the correct 1$\sigma$ errors for
a Gaussian distribution), and by using the bootstrap technique.
The results from the two methods are consistent with each other within
10\%; the values given in Table 1 are from the $S_{min} + 1$ method.
We find that the best-fit luminosity function predicts a total number of
$\int\! \int \Psi(M, z) p(M, z) \frac{dV}{dz} dM dz = 38.7 (38.6)$ 
quasars observed
in the survey area for the $\Omega=1$ ($\Lambda$) model, 
compared to the 39 quasars actually observed in the survey. 

\subsection{Non-Parametric Determination : Lynden-Bell's C$^-$ Estimator}

If the luminosity and redshift distributions of objects in a sample
are uncorrelated (\S2.2), 
the bivariate luminosity function $\Psi(L,z)$ can be separated:
$\Psi(L,z) = \psi(L)\rho(z)$,  
where $\psi(L)$ and $\rho(z)$ are the marginal distributions
of the luminosity function in the redshift and luminosity 
directions. 
In this case, the cumulative marginal distribution in the luminosity direction,
$\phi(L) \equiv \int \psi(L) dL$ can be estimated using Lynden-Bell's $C^-$ estimator
(\cite{LB71}):
\begin{equation}
\phi(L_j) = \phi(L_1) \prod_{k=2}^{j} (1+ 1/N_k),
\end{equation}
where $N_k$ is the number of objects in the comparable set of the
object $k$ as defined in Eq. (13),
and the objects are sorted according to their luminosities:
$L_1 > ... > L_{i-1} > L_{i} > ... > L_N$.
Similarly, the cumulative marginal distribution in  redshift
$\sigma(z) = \int \rho(z) dz$  can be written as:
\begin{equation}
\sigma(z_i) = \sigma(z_1) \prod_{k=2}^{i} (1+ 1/M_k),
\end{equation}
where $M_k$ is the number of objects in the box $L > L_k$ and
$z < z_{max}(L_k)$, and $z_{max}(L_k)$ is the maximum redshift the
survey can detect at the luminosity $L_k$.
Therefore, $M_k$ is the size of the comparable set in the redshift
direction, defined in a similar fashion to $N_{k}$.
Here the objects are sorted in the order of their redshifts:
$z_1 < ... < z_{i-1} < z_i < ... < z_N$.
If the $L$ and $z$ distributions are correlated, one should find a set of
parameters $x=x(L,z)$ and $y=y(L,z)$,  whose distributions
are uncorrelated, and use the $C^-$ estimator to obtain the marginal
distributions in $x$ and $y$ (e.g., Maloney \& Petrosian 1999).
Note that these equations only give the shape of the distribution, 
but not the overall normalization of the function. 
The distributions can be normalized by requiring tha the total predicted
number of objects equal that observed:
\begin{equation}
N_{obs} = \int_0^{\infty} \psi(L) \sigma[z_{max}(L)] dL =
\int_0^{\infty} \rho(z) \phi[L_{min}(z)] dz
\end{equation}

As shown in \cite{P92}, a variety of non-parametric methods can be 
reduced to Lynden-Bell's $C^-$ method (\cite{LB71}) in the limiting
case of one object per bin. The $C^-$ estimator makes efficient use
of the data by deriving the marginal distribution for the uncorrelated
variables, and does not require binning when calculating the cumulative
distribution. Obviously, binning is still needed when deriving the
differential distribution. 
But in this case, rather than binning the data in both $L$ and $z$
as for the $1/V_a$ estimator,
we only need to bin the data one axis at a time. 

For survey with a complicated selection function $p(L, z)$, the definition
of the comparable set should be changed
(Appendix A). For the luminosity direction:
\begin{equation}
N_i = \sum_j \frac{p(L_i, z_j)}{p(L_j, z_j)}, \hspace{0.5cm} \mbox{where the sum extends over} \hspace{0.2cm}
L_j > L_i,
\end{equation}
and for the redshift axis:
\begin{equation}
M_i = \sum_k \frac{p(L_k, z_i)}{p(L_k, z_k)}, \hspace{0.5cm} \mbox{where the sum extends over} \hspace{0.2cm}
z_k < z_i.
\end{equation}
With these definitions, Eqs.~(29) and (30) remain valid.

We use the equations above to derive the marginal distributions
for the SDSS sample.
For the shape of the luminosity function,
we calculate the quantity $\psi(M)/\phi(M<-25.5)$.
the number density of quasars as function of redshift, normalized
by the total number at $M<-25.5$.
For the redshift evolution, we calculate $\rho(z, M<-25.5) $, 
the total spatial density of quasars with $M_{1450} < -25.5$ as
a function of redshift normalized via Eq.~(31). 
Note that $\rho(z, M<-25.5) = \rho(z)\phi(M<-25.5)$,
therefore the product of the two quantities gives the differential
luminosity function:
these distributions are shown in Figure 2 for
both the $\Omega=1$ model and the $\Lambda$ model,
and compared with the maximum likelihood results.
The data are binned to have roughly the same number of objects per bin.
From Figure 2, we find $\psi(M_{1450}) \propto 10^{-0.55 M_{1450}}$,
and $\rho(z) \propto 10^{-0.51 z}$ for the $\Omega=1$ model.  
These two coefficients are $-0.63 \pm 0.10$ and $-0.48\pm 0.15 $
in the maximum likelihood fits (the dashed lines in Figure 2).
Comparing Figures 1 and 2, the $C^-$ results
benefit from the smaller bin size and have higher signal-to-noise ratio 
because they use marginal distributions.

\section{Evolution of the Quasar Luminosity Function at High Redshift}

In \S2, we calculated the quasar luminosity function from the SDSS
Equatorial Stripe sample using both parametric and non-parametric
methods.
This quasar sample includes 39 quasars at $i^* \lesssim 20$  and
covers 182  deg$^2$. 
Among them, 18 objects are at $z>4.0$ and six at $z>4.5$.
This is the largest complete quasar sample at $z>3.6$ to date.
We derive the luminosity function over the range
$3.6 < z < 5.0$ and $ -27.5 <  M_{1450}  < -25.5 $ (for the $\Omega = 1$
model).
The three methods (the $1/V_a$ estimator, the maximum likelihood solution
and Lynden-Bell's $C^-$ estimator) yield consistent results:

\begin{enumerate}
\item Over the redshift and luminosity range considered, 
the distribution of $z$ and $L$ are uncorrelated, therefore,
the luminosity function can be separated:
$\Psi(L,z) = \psi(L) \rho(z)$.

\item The quasar number density declines rapidly towards high redshift.
It can be fitted by an exponential decline:
$\rho(z) \sim 10^{-0.5z} \sim e^{-1.15z}$; 
the spatial density drops by a factor of
$\sim 3$ per unit redshift.

\item The quasar number density rises towards fainter luminosity as
$\psi(M) \sim 10^{0.6} M$,  or $\psi(L) \sim L^{-2.5}$.
The density increases by a factor of $\sim 4$ per magnitude.
\end{enumerate}

In this section, we first compare these results with previous
samples at high redshift, and  then discuss 
the shape of the high-redshift luminosity function.

\subsection{Comparison with Other Surveys at $z > 3.6$}

The quasar luminosity function at $z > 3.6$ has been calculated
for several previous optical quasar surveys (WHO, SSG,
DPOSS).
The selection of all these surveys is based upon the observations
in the rest-frame UV, as in our survey. 
However, some of the results are expressed in the absolute $B$-band
magnitude $M_B$ in the rest-frame.
The UV magnitudes are converted to B-band magnitudes assuming a 
power law continuum $f_\nu \propto \nu^{-0.5}$ (SSG),
which we adopt for consistency.
From Eq. (5), we have $M_B = M_{1450} - 0.48$. 
We re-write our maximum likelihood
results in the $\Omega=1$ model:
\begin{equation}
\log \Phi (z,<M_{B}) = (-6.91 \pm 0.19) -
(0.48 \pm 0.15) (z-3) + (0.63 \pm 0.10) (M_{B}+26)
\end{equation}
where $\Phi$ has units of Mpc$^{-3}$.
Since all the surveys to which we will compare
are selected from the rest-frame UV fluxes at
similar redshifts, uncertainties in the K-correction will not
affect the comparison (see also \S4.1).

SSG derive the quasar luminosity function in the redshift range
$ 2.75 < z < 4.75$ for 90 quasars at $r \lesssim 21$, selected by
their Ly$\alpha$ emission in the Palomar Transit Grism Survey.
The survey covers 61 deg$^2$, and  includes 20 quasars at $z>3.6$
and 9 quasars at $z>4.0$.
For the $\Omega=1$ model, SSG find: 
\begin{equation}
\log \Phi(z, < M_B) = -6.84 - 0.43 (z-3) + 0.75 (M_B + 26).
\end{equation}

In Figure 3, we compare the quasar evolution from the SDSS and from 
the WHO and SSG surveys.
We also show the low-redshift results from the 2dF survey (\cite{2dF}).
The new SDSS result is consistent with previous surveys.
It shows very similar redshift evolution to that of
SSG, and this trend continues towards $z\sim 5$.
The SDSS luminosity function is somewhat flatter than that of
SSG, although the difference is only at the 1-$\sigma$ level, 
and is based on rather small sample sizes.
Note, however, that the median redshift of the SDSS sample ($z\sim4$)
is considerably higher than that of SSG's ($z\sim 3.3$), which could have
an effect on the luminosity function slope (see \S3.2).

The consistency of these two surveys is important because 
they used totally different selection methods and
have very different selection functions:
SSG is a slitless spectroscopic survey whose selection is based on the
detection of emission lines in grism spectra.
It is therefore is biased towards quasars
with strong emission lines, but is not selected on the
continuum shape of the quasars. 
On the other hand, the SDSS sample selection is based on 
broad-band colors.
It depends on the strength of the emission lines rather
weakly, but is biased towards
objects with stronger continuum breaks and/or bluer continua,
and is not sensitive to red quasars ($\alpha \lesssim -1.6$,  see the
discussion in Paper III).

The WHO survey (\cite{WHO}) and the DPOSS survey (\cite{DPOSS})
used broad-band colors to select quasar candidates.
These studies are based on photographic photometry from Schmidt plates,
which typically have photometric errors $\gtrsim 0.10$ mag.
They use a very similar simulation technique to Paper III
to correct for selection effects,
although they suffer from a larger selection incompleteness ($\sim 50 \%$).
The WHO survey covers an effective area of 43 deg$^2$ down to
$m_{or} = 20$.
It contains 86 quasars at $z>2.2$, including 8 at $z>3.5$ and 2 at $z>4$.
They found that the quasar number density drops by a factor of $\sim 6$
from $z=3.3$ to $z=4.0$, 
although given the small number of objects
in their sample, the rate of evolution is poorly constrained.
The DPOSS survey includes 10 quasars at $z>4$ and $16.5 < r < 19.6$,
covering 681 deg$^2$.
They derive the quasar luminosity function at their median redshift
$z=4.35$.
In Figure 4, we compare the cumulative quasar luminosity function 
from the SDSS sample
at $z\sim 4.3$ with the results from these three previous surveys.
The results are all consistent within the error bars.
Note that although the two previous multicolor surveys
needed a much larger selection correction than the SDSS sample, 
the final corrected results agree with one another quite well.

The comparisons above show that the decline of the number density
of luminous quasars from optical surveys is unlikely to be
strongly affected by either the selection technique or the correction
for the selection functions.
In \S4.1, we discuss the effect of K-correction and extinction on
the evolution of high-redshift quasar densities.

\subsection{The Shape of the High-redshift Quasar Luminosity Function}

Using the SDSS sample, we find that at $z\sim 4$ and 
$ M_B < -26$ , the shape of
the quasar luminosity function can be fitted with a power law
\begin{equation}
\psi(L) \propto L^{-2.5 \pm 0.25},
\end{equation}
where the power law index is the average of the results from 
maximum likelihood and $C^-$ estimators.

At $z<3$, the quasar luminosity function is often fitted with a double
power law: 
\begin{equation}
\Psi(M_B, z) = \frac{\Psi^*(M_B)}
{10^{0.4[(\beta_1+1)(M_B-M_B^*(z))]} + 10^{0.4[(\beta_2+1)(M_B-M_B^*(z))]}} ,
\end{equation}
In the expression above, pure luminosity evolution is assumed, 
with  a redshift dependence
given by the evolution of the characteristic luminosity $M_B^*(z)$.
Boyle et al. (2000) find that for the 2dF sample,
the normalization of the luminosity function $\Psi^*(M_B) = 1.1 \times 10^{-6}$ Mpc$^{-3}$ mag$^{-1}$, the faint end slope $\beta_1 = -1.58$, and
the bright end slope $\beta_2 = -3.43$ (the $\Omega=1$ model).
The error bar on the slopes $\beta_1$ and $\beta_2$ are $\lesssim 0.1$. 

If we assume that the luminosity function
in Eq.~(37) applies to $z\sim4$, i.e., the shape of the luminosity
function remains the same while only the characteristic luminosity
$M_B^*(z)$ varies, we find that in order to match our results 
at $M_B \sim -26$, 
we must have $M_B^* \sim -24.7$, more than one magnitude fainter than
our survey limit.
Therefore, the SDSS results in this paper  clearly 
probe the bright end of the quasar luminosity function.

The results on the bright-end slope of the low-redshift quasar luminosity
function from various surveys appear to be quite robust: $
\phi(L) \propto L^{\beta}$, with $\beta \sim -3.4 - 3.5$
(Boyle et al. 1988, Hewett et al. 1991, Pei 1995, Boyle et al. 2000),
much steeper than our high-redshift results ($\beta \sim -2.5$).
At low redshift, the quasar number density drops by a factor of $\sim 10$
per magnitude at the bright end, while at $z\sim4$, it only drops by
a factor of 4.

We can further test the significance of this discrepancy.
We derive a new maximum likelihood solution  from the SDSS sample by  
forcing the slope of the luminosity $\beta = -3.43$ but allowing 
the normalization $\Psi_0$ and the redshift evolution $\alpha$ to vary
in the luminosity function (Eq. 27). 
This is shown as the dashed line in Figure 5.
The luminosity function from
the SDSS sample using both the $C^-$ estimator and the maximum likelihood
estimator from Figure 2 are shown as the points and solid line
in the figure.
It is evident that this steep slope is not consistent with the data.
The statistic $S$ (Eq. 22) of 
the maximum likelihood solution follows a $\chi^2$ distribution.
In this case, the likelihood ratio  for a specific $\beta_0$ 
can be estimated by calculating $\Delta S = S(\beta_0) - S_{min}$,
where $S_{min}$ is the minimum $S$ value when all three parameters are allowed
to vary, and $S(\beta_0)$ is the $S$ value if $\beta$ is fixed to
be $\beta_0$ and the other two parameters are allowed to vary.
We find that by forcing $\beta = -3.43$,
$\Delta S = 8.9$ with two degrees of freedom.
This corresponds to a probability $p = 1\%$, 
so  the two slopes are inconsistent at $\sim$ 2.5-$\sigma$ level. 

We can use another method to estimate the significance of 
the inconsistency between low and
high-redshift slopes:
there are six quasars at $M_{1450} < -27$ in the sample.
Using the model with $\beta = -3.43$, only $\sim 1.7$ quasars are predicted
over the whole survey.
The probability of $\geq 6$ quasars being observed in our survey is
only $0.8 \%$ from Possion statistics.

It is unlikely that this flattening arises  because we are missing a large
fraction of low luminosity quasars.
The photometric errors are  quite small and the selection function
is very uniform as a function of magnitude (Figure 8 of Paper III)
for all objects a few tenths of a magnitude brighter than the survey limit.
Figure 5 shows that the slope is much flatter than $\beta=-3.43$
even over  the  three most luminous magnitude bins.
Thus the bright end slope of the quasar luminosity function is considerably flatter 
at $z\sim4$  than at $z\sim 2$.
Note that SSG find $\psi(L) \propto L^{-2.9}$, while their median redshift
is somewhat lower.
The flattening of the bright end luminosity function at high-redshift, 
if confirmed  using a larger sample, indicates that the evolution
of quasar luminosity function at $z>3$ cannot be pure luminosity evolution,
unlike the case for $z<3$. Instead, the shape of the quasar luminosity
function evolves with redshift as well. 

\section{Discussion}

Figure 3 shows that the  comoving quasar number density peaks at $z\sim 2.5$, 
and drops by a factor of $\sim 20$
from $z \sim 2.5$ to  5.
This decline is evident in all optical high-redshift quasar surveys to date.
However, these results might be biased by three factors:
the survey incompleteness; uncertainties in the K-correction when comparing
low and high redshift results; and the possibility that a large fraction
of quasars are not detectable in optical survey due to dust extinction.

In Paper III, we show that the survey selection criteria are only sensitive
to a range of the quasar SED shapes, and that the incompleteness is a function
of redshift.
When calculating the luminosity function, we first correct this incompleteness
by assuming that the distribution of the continuum power-law index
$\alpha$ is a Gaussian with mean $-0.79$ and standard deviation  $0.34$.
At $4.5 < z < 5.2$, the survey is sensitive for quasars even with
 $\alpha < -2.0$, 
although no quasars with $\alpha < -1.6$ are detected in the sample.
At $z<4.5$, the survey is not sensitive to quasars with 
$\alpha \lesssim -1.6$. Therefore, the density at $z\sim 4$ could be
underestimated if there was a significant contribution
from quasars at $\alpha \sim -2.0$ that we failed to detect.
However, since we are not biased against these red quasars at the
highest redshift bin ($z>4.5$),
the strong decline in the quasar density seen at $z\sim 5$ will not
be affected by them.
Since our selection is not sensitive to very red quasars
($\alpha < -2.5$), 
if there were a large population of such objects
{\em only} at high redshift, 
the inferred density decline at high redshift
from our color-selected sample might be an overestimate. 
But as we pointed out in \S3.1, the same decline is observed in a
spectroscopically selected sample (SSG), which argues that neither
survey suffers from large incompleteness.

Low-redshift quasar luminosity functions usually are expressed
in the rest-frame B band magnitude $M_{B}$.
When comparing with high-redshift results, which are measured in
the rest-frame UV (e.g. $M_{1450}$),
a K-correction is needed.
It is customary to assume a $\alpha = -0.5$ power law continuum for
the K-correction.
An incorrect K-correction might result in a large difference in
density.
In fact, we find that the UV continua of quasars at $z\sim 4$ is better fit
with $\alpha \sim -0.8$, a value that was also found in 
the similar sized sample of $z>3$ quasars (Schneider, Schmidt \& Gunn 1991).
A change of continuum slope by 0.4 will result in a 0.2 mag difference
in the K-correction, or a 30\% difference in quasar density.
However, when we compare the observed results at $z\sim4$ with $z\sim 2$
the situation is actually better than this calculated difference
would imply.
The low-redshift quasar surveys (such as 2dF) are selected based on 
the observed $B$-band magnitude ($\lambda_{eff} \sim 4400$\AA).
At $z\sim 2$, the $B$-band corresponds to 1470 \AA\ in the rest-frame, 
almost the same as the case in which we select $z\sim 4$ quasars
based on their $i'$ band measurements ($\lambda_{eff} \sim 7600$\AA).
Therefore, both results are based on the rest-frame UV flux; 
the comparison is not strongly affected by the K-correction,
as long as the two results are transferred to the rest-frame B-band
in the same way.

The presence of dust might bias the results on quasar evolution.
Dust could arise along the line of sight in
the damped Ly$\alpha$ systems, or in the quasar environment.
\cite{FP93} studied the effect of dust in damped Ly$\alpha$ systems
on the determination of the quasar luminosity function.
They found that 10\% - 70\% of quasars at $z\sim 3$ could be 
missed from optical surveys whose selection was based on 
B band magnitudes.
This fraction cannot be calculated precisely, due to uncertainties
in the dust content and dust distribution in damped Ly$\alpha$ systems.
But the effect of intervening dust must increase  quickly
with redshift, both because the total number of damped systems 
increases, and because the extinction curve rises sharply towards
the rest-frame UV.
The number density of quasars at $z>4$ could be
significantly affected by dust along the line of sight,
but it is not clear whether it is enough to change the general trend
of the decline.

High-redshift quasars are known to reside  in dusty environments
(\cite{Omont96}, \cite{Carilli00}),
and the new millimeter and submillimeter surveys have revealed
a population of high-redshift dusty galaxies
(e.g. \cite{Barger98}).
The existence of dust extinction changes the bolometric
correction from the rest-frame UV luminosity observed in the optical
surveys.
If the bolometric correction is a strong function of redshift, 
the evolution of the bolometric luminosity function of quasars
could be very different from that of optically selected quasars.
However, we've seen that the $\alpha$ distribution of $z \approx 4$ quasars is quite
similar to that of lower-redshift quasars, implying that the objects in our sample
are no more reddened than those at low redshift.  Thus in order to invoke dust
to explain the decline of optical quasars with redshift, one would require a completely separate,
very extincted population of objects whose numbers increase with redshift.

Radio and X-ray selected samples are not strongly affected
by dust extinction.
\cite{Shaver96} found that the space density of high-redshift radio
quasars also declines rapidly at $z>3$.
On the other hand, \cite{Miyaji00} found that 
there is no evidence for a decrease in the  number density of ROSAT-selected
X-ray quasars at
$z>3$, although the statistical significance of this result is only marginal.
One difficulty is how to normalize the results from different wavelength
samples.
At $z\sim 3$, the spatial density of the quasars
in the \cite{Miyaji00} sample is considerably higher than 
in the SSG sample, which implies that these X-ray quasars are
intrinsically fainter objects than those in SSG.
The redshift evolution is expected to depend on luminosity, 
and there are theoretical reasons to believe that the evolution
is flatter at low luminosities (see below).


Studies of the quasar luminosity function put important constraints
on models of quasar evolution.
Haiman \& Loeb (1998) show that the observed quasar luminosity function can be
fit with a  theoretical model based on 
the Press-Schechter (1974) approximation, 
with reasonable assumptions about the relation between
quasar luminosity and dark matter halo mass, and the duty cycle
of quasar activity.
Haiman \& Hui (2000) and Martini \& Weinberg (2000) further demonstrate that the duty cycle
can be better constrained by fitting both the luminosity function and
two-point correlation function of high-redshift quasars. 
Faint quasars reside in less massive dark matter halos in this model.
They represent less rare peaks in the density field,
and are expected to evolve more slowly than are the bright quasars.
The luminosity function in this paper 
covers only two magnitudes in luminosity,
which is not enough to make a detailed comparison with these models. 
A deep high-redshift quasar survey is needed in order to
constrain the faint end slope of the quasar luminosity function
and to compare with theoretical models.
 
Observations of a $z=5.8$ quasar (\cite{Fan00}) demonstrates that 
the universe is already highly ionized at $z\sim 6$,
by either the UV ionizing radiation from quasars or
from star-forming galaxies.
Assuming that the high-redshift quasar luminosity function has the same shape as at
low redshift, Madau et al. (1999) estimated that the UV ionizing photons
from high-redshift quasars is not sufficient to keep the universe
ionized at $z\sim 4$. 
In this paper, we have shown that the shape of the high-redshift 
quasar luminosity function is  shallower than at low redshift for
bright quasars, implying even fewer ionizing photons from the these quasars
than assumed in Madau et al. (1999).
However, the faint quasars might contribute
more UV ionizing photons than do bright quasars at high redshift.
We currently have no knowledge of the faint end slope of the quasar luminosity
function at $z>3$. 
Therefore, a deep quasar survey is also needed in order to determine the nature
of high-redshift ionizing background.

The SDSS southern survey (see Paper III) will image the
Fall Equatorial Stripe (the same area of the sky as the sample used in
this paper) 35 -- 40 times in the five-year survey period. 
By co-adding these data, we will be able to select high-redshift 
quasar candidates 
down to $i' \sim 22.5$ at $z\sim 4$, and even fainter for
low-redshift quasars. 
In Paper III, we present 18 quasars down to $i' \sim 21$ selected
from areas that have been observed twice during the SDSS commissioning.
The high-redshift quasar luminosity function from the southern survey
will reach $M_{B} \sim -24$,
close to the traditional boundary between
quasars and Seyfert galaxies ($M_B = -23$).
This sample will give a more complete  description of the statistical properties of
high-redshift quasars.

\bigskip
The Sloan Digital Sky Survey (SDSS) is a joint project of the
University of Chicago, Fermilab, the Institute for Advanced Study, the
Japan Participation Group, The Johns Hopkins University, the
Max-Planck-Institute for Astronomy, Princeton University, the United
States Naval Observatory, and the University of Washington.  Apache
Point Observatory, site of the SDSS, is operated by the Astrophysical
Research Consortium.  Funding for the project has been provided by the
Alfred P. Sloan Foundation, the SDSS member institutions, the National
Aeronautics and Space Administration, the
National Science Foundation, the U.S. Department of Energy, and
Monbusho, Japan.
The SDSS Web site is {\tt http://www.sdss.org/}.
The Hobby-Eberly Telescope (HET) is a joint project of the University of Texas
at Austin,
the Pennsylvania State University,  Stanford University,
Ludwig-Maximillians-Universit\"at M\"unchen, and Georg-August-Universit\"at
G\"ottingen.  The HET is named in honor of its principal benefactors,
William P. Hobby and Robert E. Eberly.
XF and MAS acknowledge
support from Research Corporation, NSF grant AST96-16901, the
Princeton University Research Board, and a Porter O. Jacobus Fellowship.
DPS acknowledges support from NSF grant  AST99-00703.

\appendix
\section{Including the Selection Function in Lynden-Bell's $C^{-}$ Estimator}

Lynden-Bell's (1971) $C^{-}$ estimator calculates the marginal distribution
of quasars in luminosity or redshift by constructing a comparable set,
the luminosity and volume limited subsample for each object in the sample. 
The comparable set in luminosity is defined as:
\begin{equation}
J_i = \{ j : L_j > L_i, L_j^- < L_i\},
\end{equation}
where $L_j^-$ is the limiting luminosity at redshift $z_j$.
For a survey with a complicated selection function such as ours, 
the limiting luminosity is not well-defined: the
selection probability goes to zero smoothly at the survey limit;
elsewhere, the selection probability is a complicated function 
of $z$, $L$  and the spectral energy distribution of the object. 
In this case, the comparable set has to be redefined to include
the contribution from this selection function.
We follow the derivation of Petrosian (1992) for the
$C^-$ estimator, now 
including selection probabilities.

Here we assume that the distribution of $L$ and $z$ are not correlated,
thus the luminosity function can be written as:
\begin{equation}
\Psi(L,z) = \psi(L) \rho(z).
\end{equation}
The cumulative luminosity function is defined as
$\Phi(L,z) = \int_L^\infty \Psi(L',z) dL'$, 
\begin{eqnarray}
\frac{ d\Phi(L,z)}{\Phi(L,z)}  & = & 
  \frac{\frac{n(L,z)}{p(L,z)}dL}{\int_{L}^\infty \frac{n(L',z)}{p(L',z)}dL'} \nonumber \\
 & = & \frac{n(L,z)dL}{\int_{L}^\infty \frac{n(L',z)}{p(L',z)} p(L,z) dL'}
\end{eqnarray}
where $n(L,z)$ is the observed distribution and $p(L,z)$ is the selection function.

In order to get the marginal cumulative luminosity function,
$\phi(L) = \int_L^\infty \psi(L')dL'$, 
we need to integrate over $z$.
Since the quantity $d\phi(L)/\phi(L) = d\Phi(L,z)/\Phi(L,z)$ does not depend on $z$, 
it is easy to show that
\begin{eqnarray}
\frac{d\phi(L)}{\phi(L)}  
 & = & \frac{ \int_0^\infty dz\ n(L,z) dL}{ \int_0^\infty dz \int_L^\infty dL' \frac{n(L',z)}{p(L',z)}p(L,z)}.
\end{eqnarray}
Integrating it gives:
\begin{equation}
\phi(L) = A \exp \left\{ \int_L^{\infty} dL 
\frac{ \int_0^\infty dz\ n(L,z) dL}{ \int_0^\infty dz \int_L^\infty dL
' \frac{n(L',z)}{p(L',z)}p(L,z)} \right\},
\end{equation}
where $A$ is a constant.

In the discrete case, the observed distribution is the sum of 
a series of $\delta$ functions over individual objects:
\begin{equation}
n(L,z) = \sum_{i} \delta(L-L_i) \delta(z-z_i).
\end{equation}
Let us  assume that the objects in the sample are sorted by their luminosities:
$L_1 > L_2 > ... > L_{i-1} > L_i > ... > L_N$. 
Note that for the discrete case, $\phi(L)$ is the step function:
$\phi(L_j-\epsilon) = \phi(L_{j-1})$, and
$\phi(L_j+\epsilon) = \phi(L_{j})$, when $\epsilon \ll 1$.
Thus, one re-writes the integral from $L_j-\epsilon$ to
$L_j+\epsilon$, and regards $p(L,z)$ 
as a constant over the integrated range: 
\begin{equation}
\phi(L_j)  =  \phi(L_{j-1}) 
 \exp \int_{ L_j-\epsilon}^{L_j+\epsilon} \frac{\delta(L-L_j) dL}
{\sum_{L_i > L_j} \frac{p(L_j,z_i)}{p(L_i,z_i)} + \Theta(L-L_j)}, 
\end{equation}
where $\Theta(L)$ is a step function.
We can define the comparable set for object $j$ as:
\begin{equation}
J_j = \{ i: L_i > L_j \},
\end{equation}
and
\begin{equation}
N(L_j) = \sum_{i\in J_j} \frac{p(L_j, z_i)}{p(L_i, z_i)}.
\end{equation}
Note that $d\Theta(x) = \delta (x)$. We have:
\begin{eqnarray}
\phi(L_j)  & = & \phi(L_{j-1}) \exp \left\{ \int_{L_j-\epsilon}^{L_j+\epsilon} 
\frac{d\Theta(L-L_j)}{ N(L_j) + \Theta(L-L_j)} \right\} \\
 & = & \phi(L_{j-1}) \frac{N(L_j)+1}{N(L_j)}
\end{eqnarray} 
Thus, we have the $C^-$ estimator for the cumulative luminosity function:
\begin{equation}
\phi(L_j) = \phi(L_1) \prod_{i=2}^{j}(1+1/N_i)
\end{equation}
for the case of an arbitrary selection function.

Similarly, if we define:
\begin{equation}
M_i = \sum_j \frac{p(L_j,z_i)}{p(L_j,z_j)}, \hspace{0.2cm}
\mbox{where the sum extends over} \hspace{0.1cm}  z_j < z_i,
\end{equation}
and the sample is sorted by redshift,
$z_1 < ... < z_{j-1} < z_j < ... < z_N$,
the cumulative redshift distribution can be estimated as:
\begin{equation}
\sigma(z_k) = \sigma(z_1) \prod_{i=2}^{k} (1+1/M_i).
\end{equation}

\newpage
\begin{deluxetable}{crr}
\tablenum{1}
\tablecolumns{3}
\tablecaption{Maximum Likelihood Solutions}
\tablehead
{
Parameter & $\Omega=1$  & $\Omega=0.35$  $\Lambda=0.65$ \\
	  & $h=0.5$ & $h=0.65$
}
\startdata
Differential, Eq. (26) & & \\ \hline \\
\smallskip
$\Psi_0$ (Mpc$^{-3}$ mag$^{-1})$ &  $8.4^{+4.6}_{-3.1} \times 10^{-8}$ &$ 7.2^{+4.0}_{-2.6} \times 10^{-8}$ \\ 
\smallskip
$\alpha$ & $0.76 \pm 0.29 $ & $ 0.75 \pm 0.29$ \\
\smallskip
$\beta$ & $-2.58 \pm 0.23$ & $-2.58 \pm 0.23$ \\ \hline 
\smallskip
 Cumulative, Eq. (25) & & \\ \hline \\
\smallskip
A & $-7.24 \pm 0.19$ & $-7.31 \pm 0.19$ \\
\smallskip
B & $0.48\pm 0.15$ & $0.47 \pm 0.15$ \\
\smallskip
C & $0.63 \pm 0.10$ & $0.63 \pm 0.10$
\enddata
\end{deluxetable}

\newpage
\begin{figure}
\vspace{-3cm}

\epsfysize=600pt \epsfbox{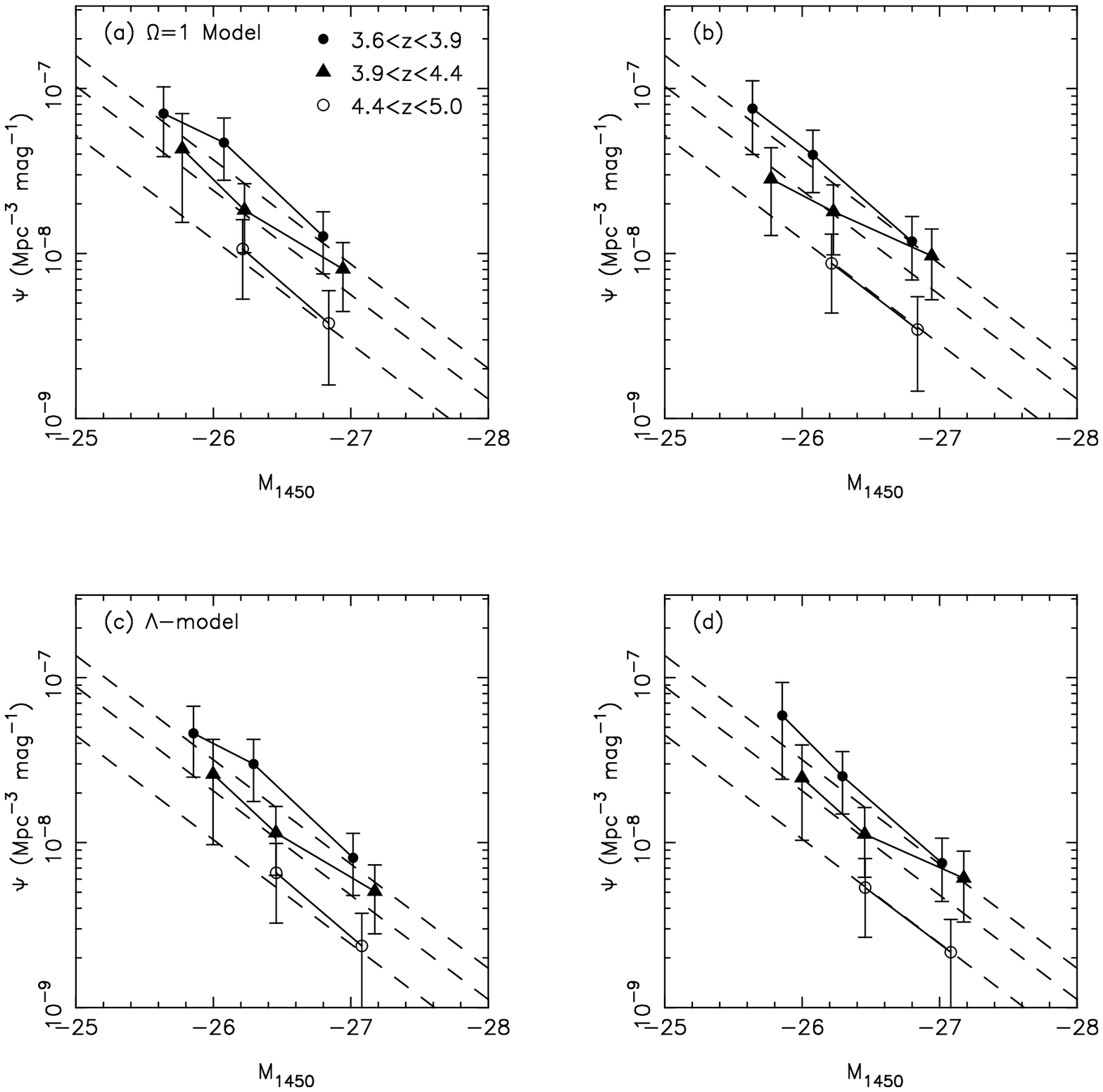}

\vspace{1cm}
Figure 1. The high-redshift quasar luminosity function derived from
the $1/V_a$ estimator. 
(a) is the result for the $\Omega=1$ model, corrected for the average selection function;
(b) is the result for the $\Omega=1$ model, corrected for the selection function using 
each quasar's SED type;
(c) and (d) are for the $\Lambda$-model with the two methods of selection function
correction.
The dashed lines are the maximum likelihood solutions.
\end{figure}

\begin{figure}
\vspace{-3cm}

\epsfysize=600pt \epsfbox{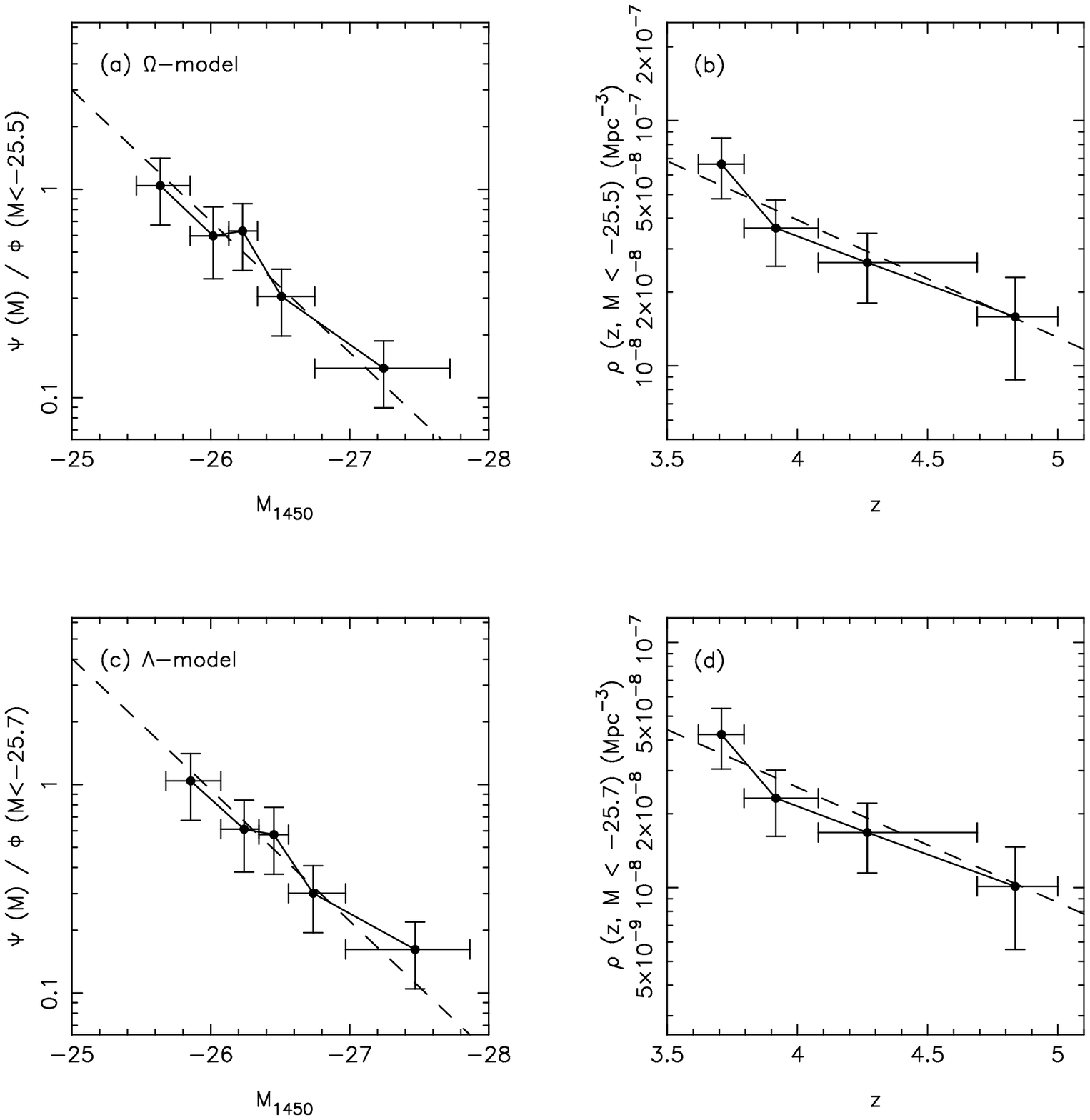}
\vspace{1cm}
Figure 2. The high-redshift quasar luminosity function derived using
Lynden-Bell's (1971) $C^{-}$ estimator.
(a) and (c) show the marginal differential luminosity distribution 
as a function of magnitude, for both the $\Omega=1$ and the $\Lambda$-model;
(b) and (d) show the marginal redshift evolution:
the spatial density of quasars at $M_{1450} < -25.5$ as a function of
redshift.
The dashed lines are the maximum likelihood solutions.
\end{figure}

\begin{figure}
\vspace{-3cm}

\epsfysize=600pt \epsfbox{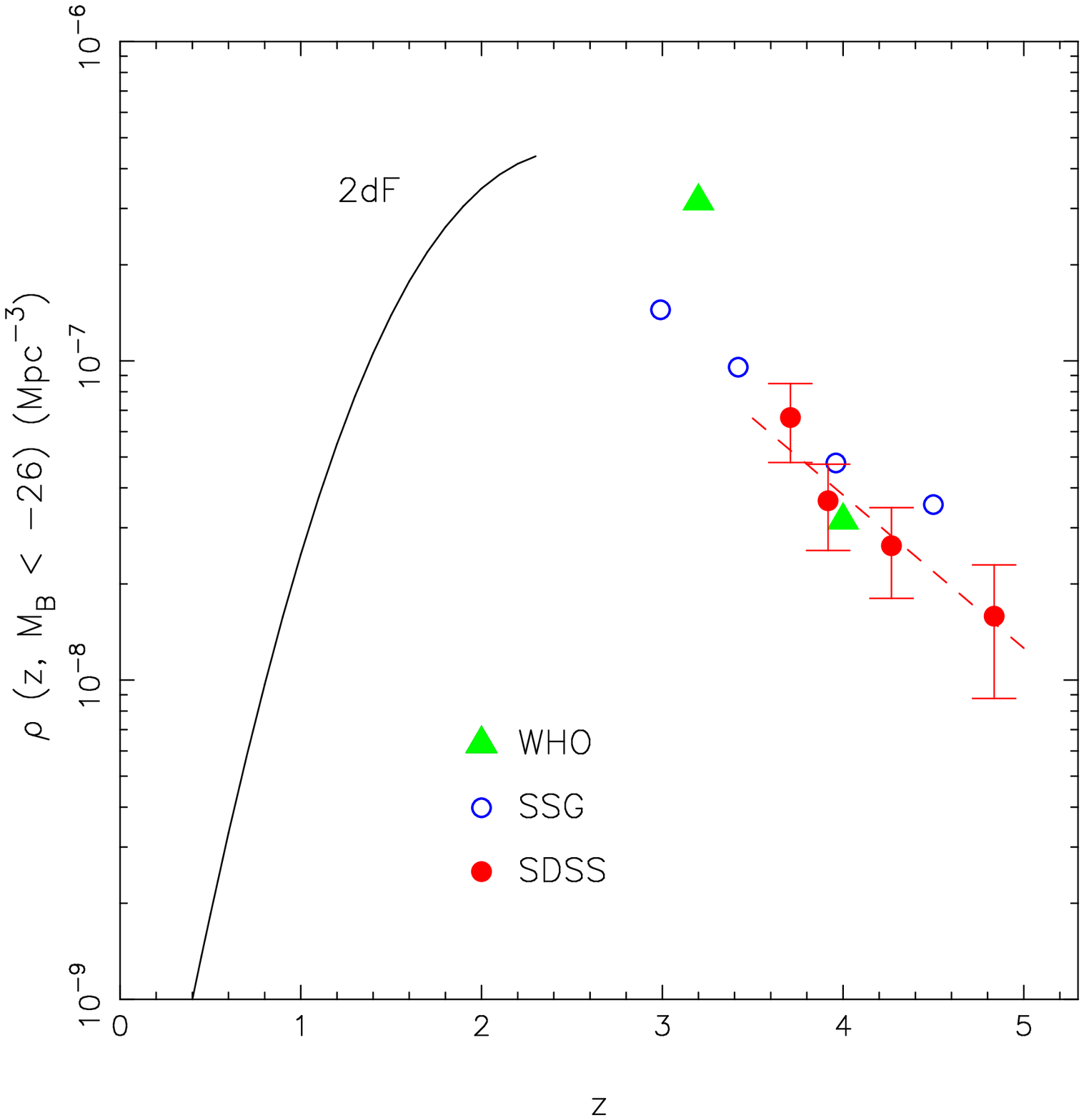}
\vspace{1cm}
Figure 3. The evolution of the quasar spatial density at $M_{B} < -26$
is compared with previous studies.
The SDSS points shown are the results using the $C^-$ estimator,
and the dashed line is the maximum likelihood solution.
The SSG points shown are the spatial density calculated with 
the $1/V_a$ method. The low-redshift result is the best-fit model
from  the 2dF survey (Boyle et al. 2000).
\end{figure}

\begin{figure}
\vspace{-3cm}

\epsfysize=600pt \epsfbox{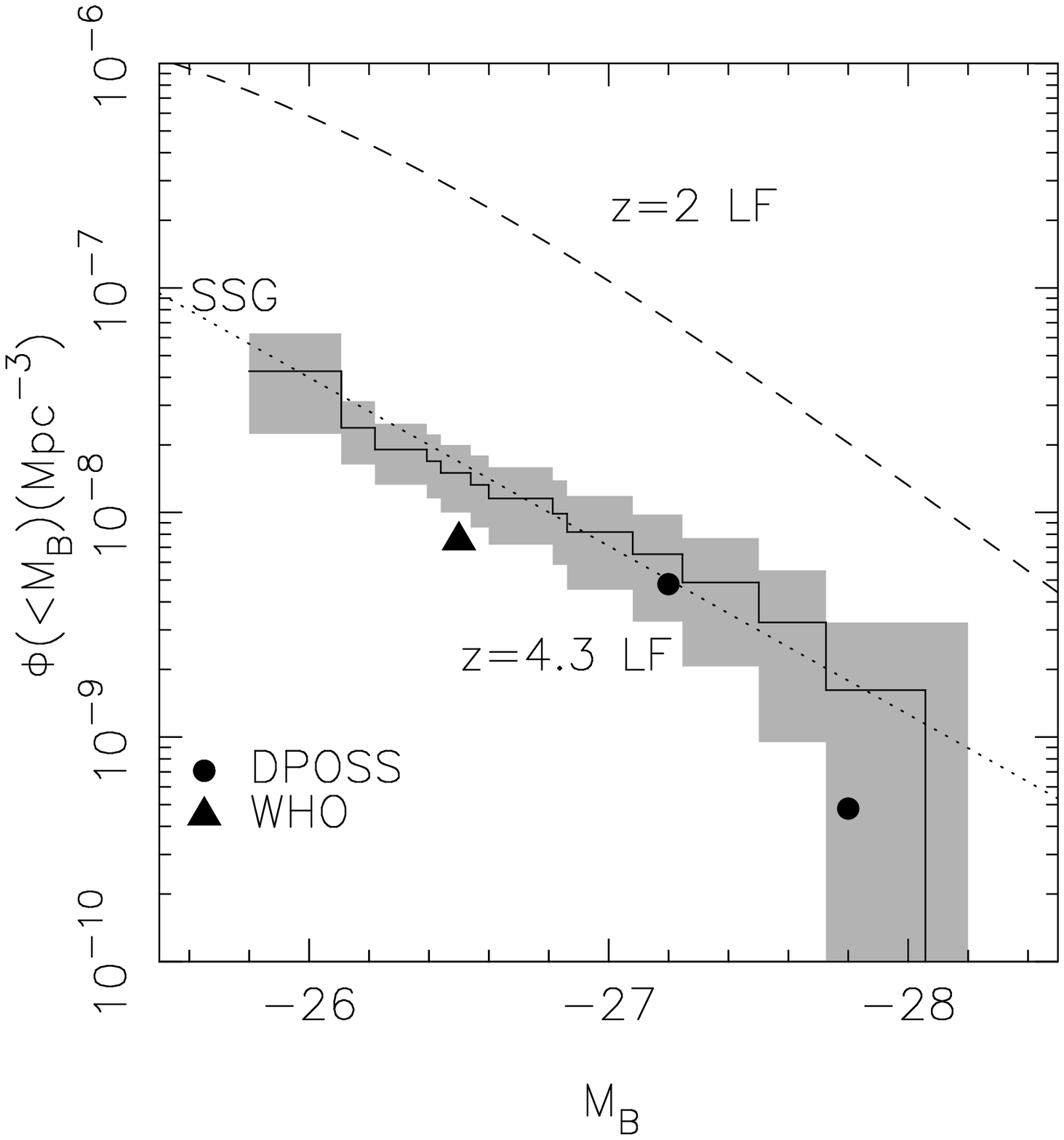}
\vspace{1cm}
Figure 4. The cumulative luminosity function at $z\sim 4.3$,
compared with previous studies.
The SDSS result  shown (shaded area) is based on the $1/V_a$ estimator for
$4.0 < z < 4.5$.
The SSG result is their best-fit model (Eq. 35).
For the $z=2$ luminosity function, we use the  best-fit model
from  the 2dF survey (Boyle et al. 2000).
\end{figure}

\begin{figure}
\vspace{-3cm}

\epsfysize=600pt \epsfbox{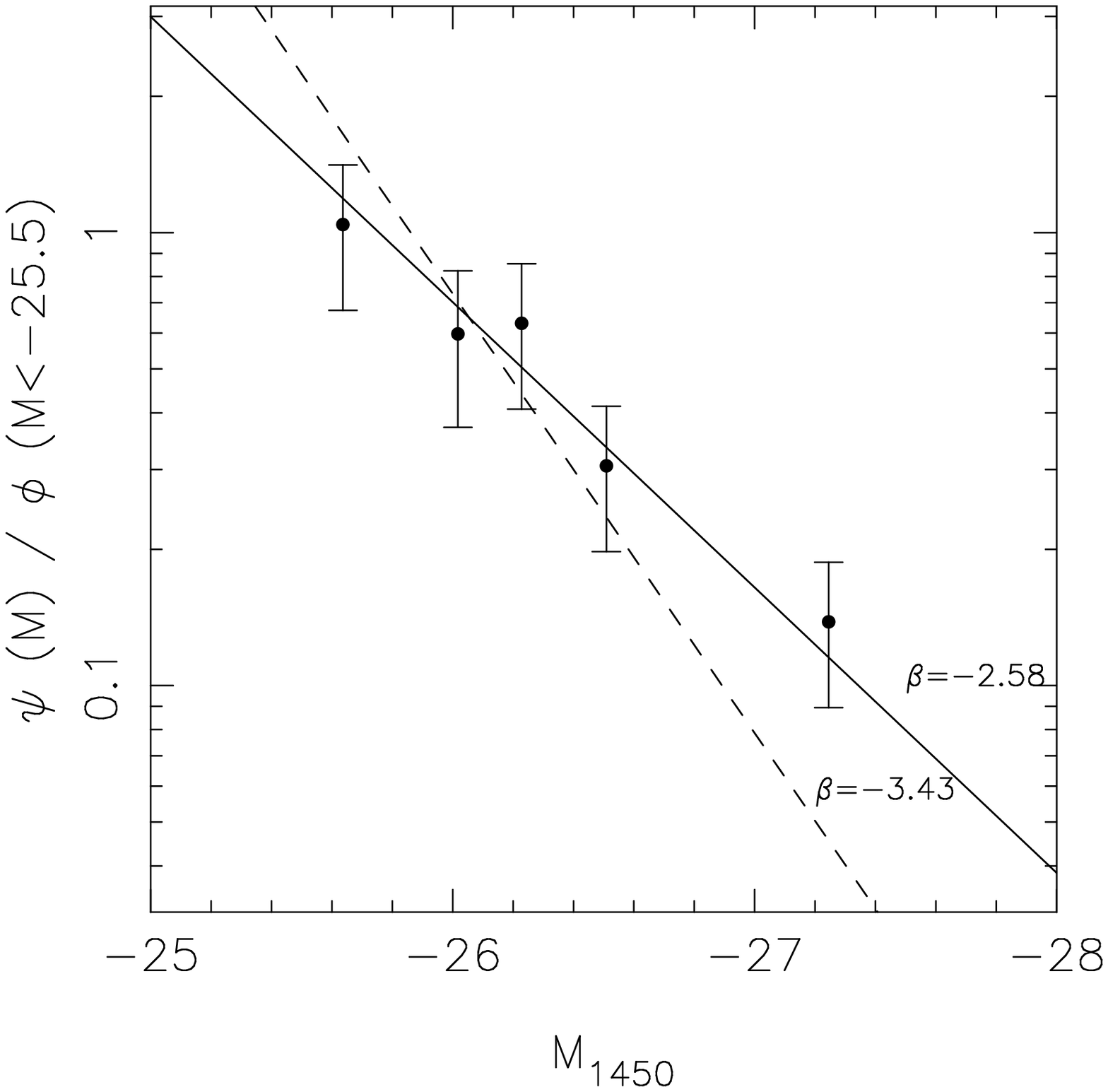}
\vspace{1cm}
Figure 5. The slope of the high-redshift quasar luminosity function
at the bright end.
The SDSS result gives $\psi(L) \propto L^{-2.58}$ (solid line), considerably shallower than
the low-redshift luminosity function $\psi(L) \propto L^{-3.43}$.
The dashed line in the plot is the maximum likelihood solution
from the SDSS sample if the slope of the luminosity function
is forced to have $\beta = -3.43$.
   
\end{figure}
\end{document}